\documentclass[aps,prl,amsmath,twocolumn,showpacs]{revtex4-2}
\usepackage{natbib}
\usepackage[english]{babel}
\usepackage{amsfonts,amsmath,amssymb,bm}
\usepackage[fleqn,tbtags]{mathtools}
\usepackage{graphicx}
\usepackage[usenames,dvipsnames]{xcolor}
\usepackage{graphicx,psfrag,dsfont,amsmath}
\usepackage{slashed,amssymb,color,pifont}
\usepackage[dvipsnames]{xcolor}
\usepackage{ulem}

\usepackage{array}
\usepackage{amsthm}
\usepackage{amsmath}
\usepackage{amssymb}
\usepackage{slashed}
\usepackage{bm}
\usepackage{graphicx}
\usepackage{wrapfig}
\usepackage{physics}
\usepackage{mathtools}
\usepackage{subfiles}
\usepackage{color, colortbl}
\usepackage{hyperref}
\usepackage{marginnote}
\usepackage{boxhandler}
\usepackage{xr}
\usepackage{float}

\newcommand{\bea}{\begin{eqnarray}}
\newcommand{\ena}{\end{eqnarray}}

\date{}

\begin{document}
\title{The integer quantum Hall transition: an $S$-matrix approach to random networks}

\author{H. Topchyan$^{1}$, I. Gruzberg$^{2}$, W. Nuding$^{3}$, A. Kl\"umper$^{3}$  and A. Sedrakyan$^{1}$}
\affiliation{$^1$Yerevan Physics Institute, Br. Alikhanian 2, Yerevan 36, Armenia 
\\	
$^2$Ohio State University, Department of Physics, 191 West Woodruff Ave, 
Columbus OH, 43210
\\
$^3$Wuppertal University, Gau\ss stra\ss e 20, 42119 Wuppertal, Germany}
	
\begin{abstract}	

In this paper we propose a new $S$-matrix approach to numerical simulations of network models and apply it to random networks that we proposed in a previous work~\cite{Gruzberg-Geometrically-2017}. Random networks are modifications of the Chalker-Coddington (CC) model for the integer quantum Hall transition that more faithfully capture the physics of electrons moving in a strong magnetic field and a smooth disorder potential. The new method has considerable advantages compared to the transfer matrix approach, and gives the value $\nu \approx 2.4$ for the critical exponent of the localization length in a random network. This finding confirms our previous result and is surprisingly close to the experimental value $\nu_{\text{exp}} \approx 2.38$ observed at the integer quantum Hall transition but substantially different from the CC value $\nu_\text{CC} \approx 2.6$. 

\end{abstract}
	
\date{{\today}}
	
\maketitle

\textit{Introduction.} The integer quantum Hall (IQH) transition~\cite{Huckestein-Scaling-1995} is the best studied example of an Anderson localization-delocalization transition~\cite{Evers-Anderson-2008}. In spite of all the efforts over the years, understanding the IQH plateau transitions remains an important problem of modern condensed matter physics. Numerous experiments~\cite{Wei-Experiments-1988, Koch-Experiments-1991, Koch-Size-dependent-1991, Koch-Experimental-1992, Engel-Microwave-1993, Wei-Current-1994, Li-Scaling-2005, Li-Scaling-2009, Giesbers-Scaling-2009} have provided evidence of scaling behavior near the IQH transition, characterized, in particular, by the critical exponent $\nu$ that describes the divergence of the localization length at the transition. Over the years, the experimental value of $\nu_{\text{exp}} \approx 2.4$ has been consistently observed in many systems. A very thorough study of the IQH transition in GaAs/AlGaAs heterostructures~\cite{Li-Scaling-2005, Li-Scaling-2009} gave the value of $\nu_{\text{exp}} \approx 2.38\pm 0.06$, albeit with an important caveat (discussed in~\cite{Pruisken-Comment-2009}).

On the other hand, most numerical studies of the IQH transition in the past fifteen years reported results in the range $\nu \sim 2.5$--$2.6$, see Refs.~\cite{Slevin-Critical-2009, Obuse-Conformal-2010, Amado-Numerical-2011, Dahlhaus-Quantum-2011, Fulga-Topological-2011, Obuse-Finite-2012, Slevin-Finite-2012, Nuding-Localization-2015, Ippoliti-Integer-2018, Zhu-Localization-length-2019, Puschmann-Integer-2019, Kluemper-Random-2019, Ippoliti-Dimensional-2020, Puschmann-Edge-state-2021, Puschmann-Green's-2021, Sbierski-Criticality-2021, Dresselhaus-Numerical-2021, Huang-Numerical-2021, Dresselhaus-Scaling-2022, Bera-Quantum-2024}. Many of these references numerically simulated the celebrated Chalker-Coddington (CC) network model~\cite{Chalker-Percolation-1988, Kramer-Random-2005} which is based on the semiclassical picture of electrons drifting along the equipotential lines of a smooth disorder potential. Tunneling across saddle points of the potential leads to hybridization of the localized states and a possible delocalization. In the CC model this picture is drastically simplified, and all scattering nodes are placed at the vertices of a square lattice. It is the regular geometry of the CC model that facilitates the application of numerical transfer matrix (TM) techniques~\cite{MacKinnon-One-Parameter-1981, MacKinnon-The-scaling-1983, Kramer-Random-2005}.

A likely source of the discrepancy between the experimental and numerical values of $\nu$ is
electron-electron interaction whose effect on the scaling near the IQH transition has been studied in Refs. \cite{Lee-Effects-1996, Wang-Short-range-2000, Burmistrov-Wave-2011, Kumar-Interaction-2022}. It was shown there that short-range interactions are irrelevant at the IQH critical point and should not modify the value of $\nu$ while the Coulomb interaction present in experimental systems is relevant. This issue is not fully understood, and the fate of the critical fixed point dominated by the Coulomb interaction remains unresolved.

\begin{figure}[t]
\includegraphics[height=3.7cm]{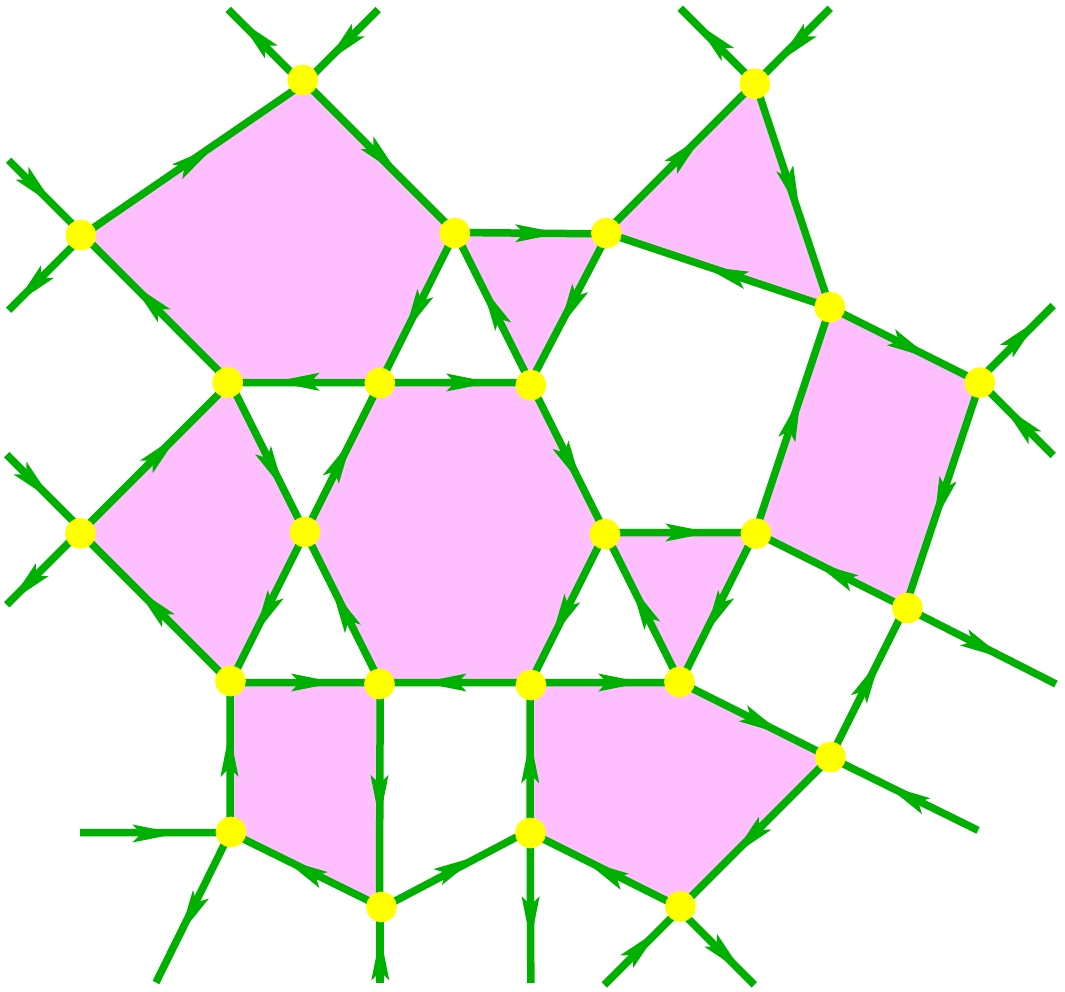}
\hfill
\includegraphics[height=3.7cm]{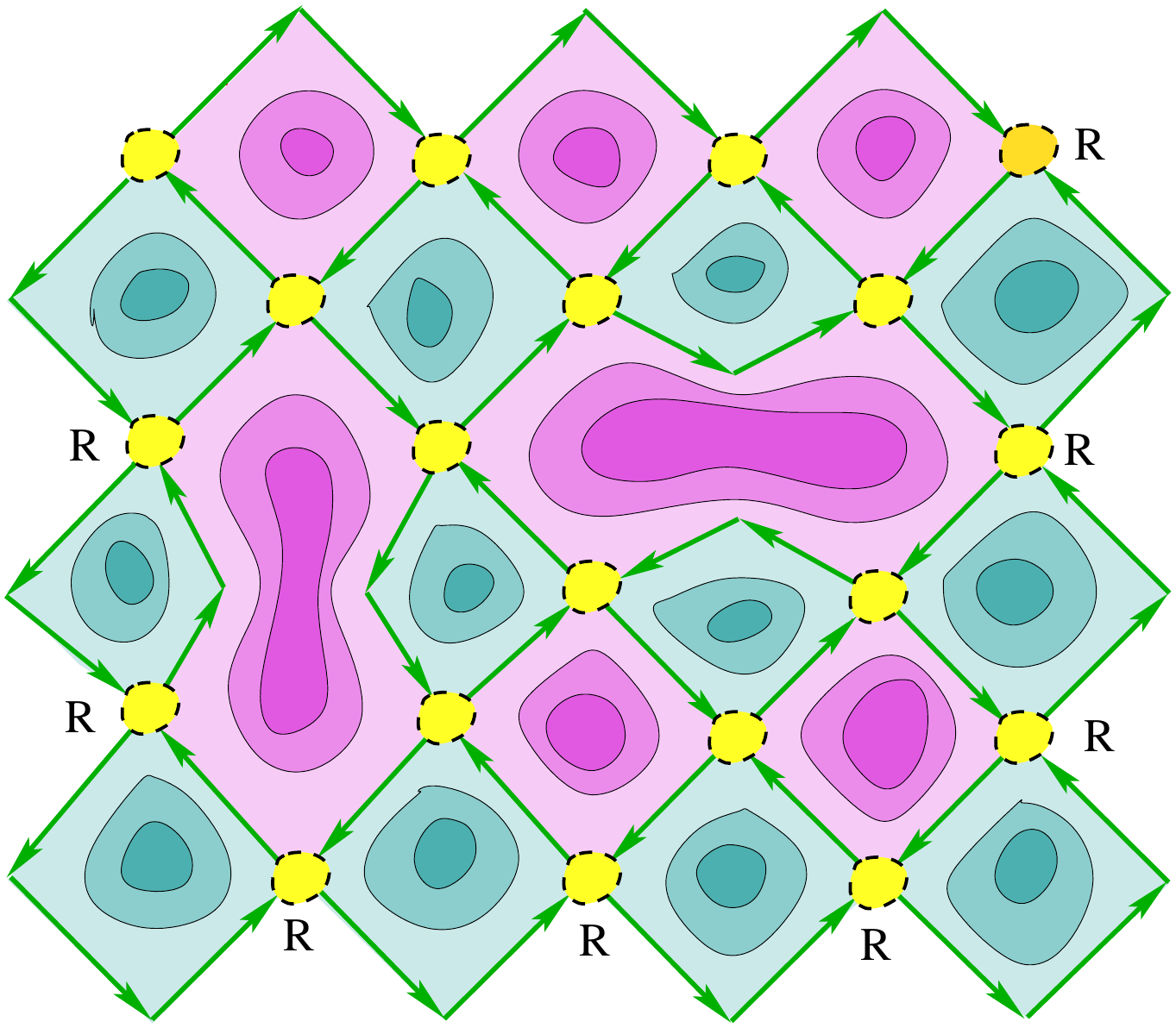}
\vskip 3mm
\includegraphics[width=0.8\linewidth]{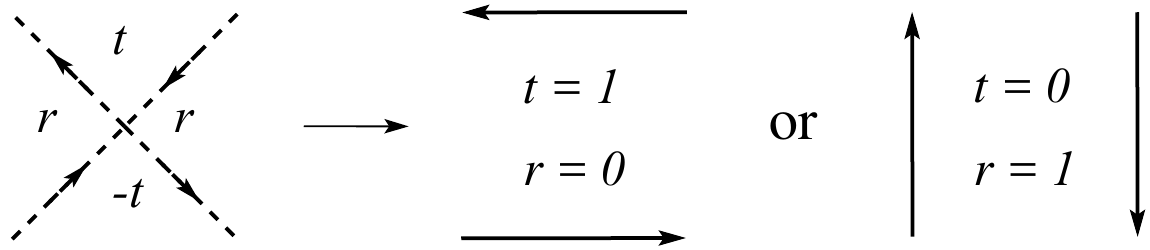}
\caption{Top left: A random network. Top right: Modified CC network with two open nodes, one in the vertical and one in the horizontal direction. Bottom: the two ways to open a node. }
\label{fig:random-graph}
\end{figure}

In Refs.~\cite{Gruzberg-Geometrically-2017, Kluemper-Random-2019} we proposed a mechanism that leads to a modification of $\nu$ from its CC value even within the single-particle framework. At the heart of this proposal is a modification of the CC model that is expected to better capture the geometric disorder inherent in the semiclassical network of drifting electron orbits. Indeed, saddle points that connect the ``puddles'' of filled electron states do not form a regular lattice, and around each ``puddle'' there may be any number of them. Taking this into account led us to consider structurally disordered, or {\it random networks} (RNs), see top left in Fig.~\ref{fig:random-graph}. Each node of the network represents a $2 \times 2$ scattering matrix 
\begin{align}
s = \begin{pmatrix*}[r] r & t \\ -t & r \end{pmatrix*}, 
\label{s-matrix}
\end{align}
and each link carries a random U(1) phase. The transmission and reflection amplitudes satisfy $t^2 + r^2 = 1$ and can be parametrized as $r = (1 + e^{-2x})^{-1/2}$, $t = (1 + e^{2x})^{-1/2}$ with $x \in [-\infty, \infty]$. At the critical point describing the IQH transition we have $x_c = 0$ and $r_c = t_c = 1/\sqrt{2}$. 

To simulate RNs numerically, we adopted the following construction. Starting with the regular CC network, at each node we set $t=1$ with probability $p \in [0,1/2]$,  $t=0$ with the same probability $p$, and left the node unchanged (with the same value of $x$ close to $x_c = 0$) with probability $1 - 2p$. The modified nodes with $t=1$ ($t=0$) are ``open'' in the horizontal (vertical) direction, and opening a node changes the four adjacent square faces into two triangles and one hexagon, see top right and bottom in Fig.~\ref{fig:random-graph}. Repeated opening of nodes can produce tilings of the plane by polygons with arbitrary numbers of edges, corresponding to a distribution of saddle points in a realistic random potential~\cite{Conti-Geometry-2021}.

This construction allowed us to use the TM method but it suffered a difficulty: $t$ and $r$ appear in the denominators of the matrix elements of TMs. Setting them to zero is a singular procedure related to the disappearance of two horizontal channels upon opening a node in the vertical direction. To avoid such singularities, we had to use $t$ or $r = \varepsilon$ with $\varepsilon = 10^{-6} - 10^{-7}$ instead of zero in the TMs. This resulted in the values of $\nu$ that varied with $p$ but were insensitive to the ad-hoc small parameter $\varepsilon$. In particular, for $p=0$ (the regular CC model) we obtained $\nu \approx 2.57$, consistent with other results for the CC model. However, for $p=1/3$ we obtained $\nu \approx 2.37$, which is surprisingly close to the experimental value $\nu_{\text{exp}} \approx 2.38$.  

In this paper, we introduce a novel $S$-matrix approach to the numerical simulation of RNs. This approach completely avoids the introduction of the small regularizing parameter $\varepsilon$. It has another advantage: it avoids the appearance of large numbers in the TMs, leading to a significant increase in the speed of numerical calculations. Consequently, we are able to analyze significantly larger network sizes.

In our analysis, we conducted simulations for $p = 0$ corresponding to the CC model and $p = 1/3$ corresponding to our model~\cite{Gruzberg-Geometrically-2017}. The results are
\begin{align}
\nu &= 2.554 \pm 0.018, & \text{ for } p &= 0,
\label{p=0-exponents}
\\
\nu &= 2.398 \pm 0.006, & \text{ for } p &= 1/3.
\label{p=1/3-exponents}
\end{align}
For both values of $p$, our current results confirm the previous results obtained with the regularized TM approach.

The results~\eqref{p=0-exponents} and~\eqref{p=1/3-exponents} violate the Harris criterion which states that bond disorder cannot change the critical behavior of a clean system (on a regular lattice) if $d \nu > 2$~\cite{Harris-Effect-1974}. Ref.~\cite{Janke-Harris-Luck-2004} argued that it should be modified in the case of random lattices. 
Our results indicate that the structural disorder introduced by opening network nodes is indeed relevant.

\begin{figure}[t]
       \centering
        \includegraphics[width=.65\linewidth]{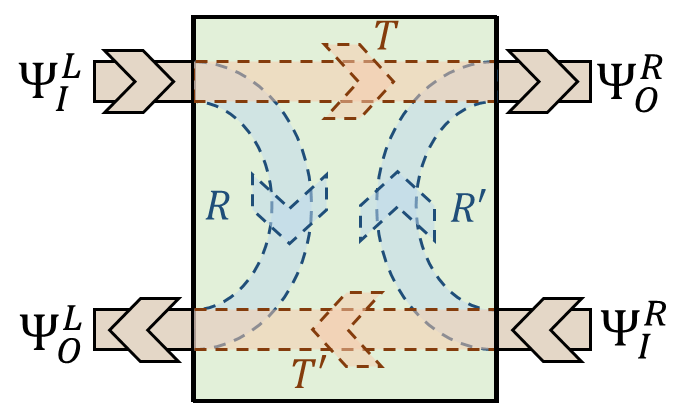}
        \caption{A single $S$-matrix structure.}
        \label{fig:s_scat}
\end{figure}

{\it S-matrix approach.} Let us remind general aspects of the scattering approach to transport in quasi-one-di\-men\-si\-onal (quasi-1D) non-interacting systems, see Ref.~\cite{Beenakker-Random-matrix-1997} for a review. We pass from the single-particle Hamiltonian to a scattering description by considering waves of a given energy that enter the system at either end, are scattered and then emerge at the same or the opposite end. We will assume that only a finite number $M$ of scattering states (or channels) in each direction are relevant at the energy considered, and denote the incoming and outgoing waves from the left and right by $\Psi^{L/R}_{I/O}$. The scattering can then be represented by a unitary $S$ matrix that maps incoming to outgoing waves:
\begin{align}
\label{S-matrix}
\begin{pmatrix} \Psi^{L}_{O} \\ \Psi^{R}_{O} \end{pmatrix}  =
S \begin{pmatrix} \Psi^{L}_{I} \\ \Psi^{R}_{I} \end{pmatrix} =
\begin{pmatrix} R  & T' \\ T & R' \end{pmatrix}
\begin{pmatrix} \Psi^{L}_{I} \\ \Psi^{R}_{I} \end{pmatrix}.
\end{align}
The four $M \times M$ blocks of the $S$ matrix represent reflection and transmission matrices, as depicted in Fig.~\ref{fig:s_scat}.
 
Alternatively, the scattering can be represented by a TM $\mathcal{T}$ that maps the in- and outgoing waves at one end (say, the left) to those at the other (the right):
\begin{align}
\label{T-matrix}
\begin{pmatrix} \Psi^{R}_{O} \\ \Psi^{R}_{I} \end{pmatrix}  =
\mathcal{T} \begin{pmatrix} \Psi^{L}_{I} \\ \Psi^{L}_{O} \end{pmatrix} =
\begin{pmatrix} A  & B \\ C  & D \end{pmatrix}
\begin{pmatrix} \Psi^{L}_{I} \\ \Psi^{L}_{O} \end{pmatrix}.
\end{align}
The advantage of the TM is that 1D systems can be composed end-to-end by multiplying elementary TMs.

Once the TM $\mathcal{T}$ of a quasi-1D system composed of $L \gg 1$ segments is found, all transport properties of the system can be obtained in terms of the $2M$ eigenvalues of $\mathcal{T}^\dagger \mathcal{T}$ which come in inverse pairs and are commonly denoted by $e^{\pm 2L\gamma_n}$ with $\gamma_n \geq 0$ ($n = 1,2,\ldots, M$). It is known that in the limit $L \to \infty$ the quantities $\gamma_n$ (called the Lyapunov exponents) are self-averaging and tend to non-random values~\cite{Oseledec-A-Multiplicative-1968}. The smallest Lyapunov exponent $\gamma_1$ determines the localization length of the quasi-1D system $\xi = 1/\gamma_1$.   

A disadvantage of the TM method is that for $L \gg 1$ the eigenvalues $e^{\pm 2L\gamma_n}$ are exponentially large and small in $L$. Dealing with such exponentially large or small numbers in simulations requires special numerically ``expensive'' methods such as the QR or LU decompositions.  

Our novel $S$-matrix approach provides a convincing workaround for both aforementioned issues: it avoids the use of a small ad-hoc parameter $\varepsilon$ as well as the appearance of large numbers in TMs. Instead, it directly utilizes $S$ matrices. The $M \times M$ blocks of the $S$ and $\mathcal{T}$ matrices are related as follows: 
\begin{align}
\mathcal{T} &= \begin{pmatrix} T - R' T^{\prime -1} R & R' T^{\prime -1} \\
-T^{\prime -1} R & T^{\prime -1} \end{pmatrix},
\label{S-to-T-matrix}
\\
S &= \begin{pmatrix} - D^{-1} C  & D^{-1} \\ 
A - B D^{-1} C  & B D^{-1} \end{pmatrix}.
\label{T-to-S-matrix}
\end{align}

\begin{figure}[t]
       \centering
        \includegraphics[width=\linewidth]{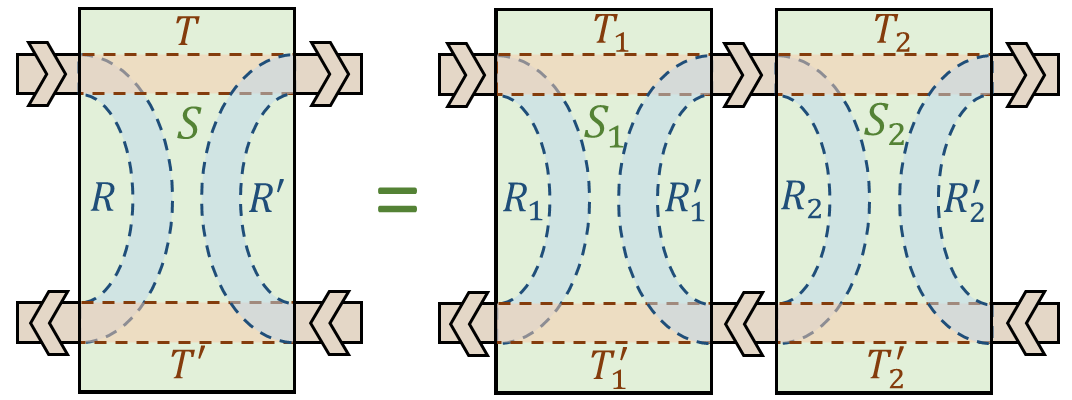}
        \caption{Combination of S-matrices. }
        \label{fig:two-scatterers}
\end{figure}

When two scatterers are composed as in Fig.~\ref{fig:two-scatterers} we can multiply their TM-s $\mathcal{T} = \mathcal{T}_2 \mathcal{T}_1$ and then convert the resulting TM to the combined $S$ matrix that is denoted as $S \equiv S_1 \star S_2$~\cite{Redheffer-on-the-relation-1962, Avishai-Quantum-1992}. The result (that defines the star product) is
\begin{equation}
\begin{split}
R &= R_1 + T_1'(1 - R_2 R_1')^{-1} R_2 T_1, \\
T &= T_2 (1 - R_1' R_2)^{-1} T_1, \\
R' &= R_2' + T_2 (1 - R_1' R_2)^{-1} R_1' T_2', \\
T' &= T_1'(1 - R_2 R_1')^{-1} T_2', 
\label{star-product}
\end{split}
\end{equation}
where the indices $i = 1,2$ denote blocks of the $S$-matrices of the two scatterers. 

The star product defined by Eq.~\eqref{star-product} is not singular when the two scatterers contain perfectly reflecting channels. Even in the extreme case when $R_2 R_1' = 1$ the corresponding transmission matrices vanish: $T_1 = T_1' = T_2 = T_2' = 0$, and the star product reduces to 
$R = R_1$, $T' = T = 0$, $R' = R_2'$, as can be seen directly from Fig.~\ref{fig:two-scatterers}. The presence of random phases on the links makes the matrices $1 - R_2 R_1'$ invertible even when there are perfectly reflecting channels. 

The star product preserves the unitarity of the scattering matrix. As a consequence, the four Hermitian matrices $T^\dagger T$, $T^{\prime \dagger} T'$, $1 - R^\dagger R $, and $1 - R^{\prime \dagger} R'$ have the same set of eigenvalues $t_1 ,t_2 , \ldots ,t_M$. Each of these $M$ transmission eigenvalues is a real number between 0 and 1 related to the Lyapunov exponents as $t_n = \cosh^{-2}(L\gamma_n)$. Thus, we can employ the star product many times without encountering exponentially large matrix elements. 

We conclude that using the composition of scattering matrices indeed solves the two problems of the TM approach that we mentioned above.  

{\it Numerical procedure.} The advantage of the $S$-matrix method that avoids exponentially large numbers leads to a significant speed-up in its numerical implementation. Indeed, as we mentioned, the TM method requires the use of QR or LU matrix decompositions which have a theoretical computational complexity that grows as $O(M^3)$ with the matrix size $M$. On the other hand, the costliest operations in the $S$-matrix method are matrix multiplications and inversions. For these operations, there are efficient algorithms whose theoretical computational complexity is $O(M^{2.38})$, substantially smaller than the naive $O(M^3)$ of brute-force basic algorithms. Thus we expect that the complexity of the $S$-matrix approach calculation is $O(M^{2.38})$ while the TM approach which also requires decompositions should have the complexity $O(M^3)$.

This substantial difference can be verified numerically through time measurements of simulations with various matrix sizes $M$. The results of our simulations indicate the complexity of $O(M^{2.18})$ for the $S$-matrix approach and $O(M^{2.79})$ for the TM approach. A probable source of deviations from theoretical values is the sparse nature of the matrices used in our simulations.

\begin{figure}[t]
\centering
\includegraphics[width=0.7\linewidth]{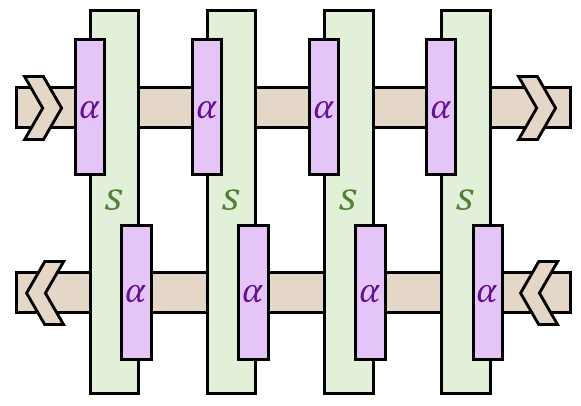}
\caption{The overall scattering structure with pure scatterings (green blocks) and random phases (purple blocks).}
\label{fig:s_mat-alpha}
\end{figure}

In practice, the $S$ matrices of the individual slices of the RN are composed of real $2 \times 2$ blocks of the form~\eqref{s-matrix}. Then the real-valued $S$ matrices are multiplied on the right by diagonal matrices $\text{diag}\,(e^{i\alpha_i})$ that represent random phases on all incoming channels. This defines the basic building block of the chain of scatterers, as illustrated in Fig.~\ref{fig:s_mat-alpha}. 
Then we compute the $L$-fold star product for a chain of $S$ matrices, and the largest eigenvalue $t_1$ of $T^\dagger T$ determines the localization length $\xi$. 

It is known~\cite{Beenakker-Random-matrix-1997} that in very long systems ($L \gg \xi$) the transmission eigenvalues $t_n$ are widely separated: $1 \gg t_1 \gg t_2 \gg \ldots \gg t_M$. In our simulations reported below, the ratio $t_2/t_1 \leq 10^{-12}$. Then, instead of diagonalizing $T^\dagger T$, we can compute its trace,
\begin{align}
\text{tr} (T^\dagger T) = \textstyle\sum_{n=1}^{M} t_n \approx t_1 \approx 4 e^{-2L\gamma_1},
\end{align}
and find an approximation $\gamma$ to the smallest Lyapunov exponent as
\begin{align}
\gamma &= - \ln[\text{tr} (T^\dagger T)/4]/2L \approx \gamma_1.  
\label{gamma}
\end{align}

This procedure is done for multiple values of the parameter $x$ close to the critical point $x_c =0$, and then the rescaled Lyapunov exponent $\Gamma(x) = M\gamma(x)$ is fit to a finite-size scaling form 
\begin{align}
M\gamma(x) = \Gamma[M^{1/\nu} u_0(x), M^y u_1(x)], 
\end{align}  
which contains one relevant and one irrelevant scaling variables $u_0$ and $u_1$. The fitting
produces the values of $\nu$ in Eqs.~\eqref{p=0-exponents} and~\eqref{p=1/3-exponents}. In addition, we obtain the irrelevant exponents $y$ and the fixed-point values $\Gamma_c = \pi(\alpha_0 - 2)$ related to the multifractal exponent $\alpha_0$~\cite{Obuse-Conformal-2010}:
\begin{align}
y &= -0.61 \pm 0.07, \,\,\, \Gamma_c = 0.816 \pm 0.0015, \,\,\, (p = 0), 
\\
y &= -0.33 \pm 0.01, \,\,\, \Gamma_c = 0.825 \pm 0.0015, \,\,\, (p = \frac{1}{3}). 
\end{align}

{\it Numerical details.} We performed numerical simulations of RNs using the $S$-matrix method for $p=0$ (CC model) and $p=1/3$ studied in Ref.~\cite{Gruzberg-Geometrically-2017}, but with larger system sizes, more values of the parameter $x$, and larger statistical ensembles. The calculations were conducted for $M$ ranging from 40 to 300 and the product length $L = 5 \times 10^6$. Following the method of Ref.~\cite{Amado-Numerical-2011}, we used ensembles containing approximately $N_r = 400$ and $N_r = 1500$ disorder samples for $p=0$ and $p=1/3$ respectively. The use of ensembles is equivalent to simulating systems with effective lengths of $L_\text{eff} = 2 \times 10^9$ for $p=0$ and $L_\text{eff} = 7.5 \times 10^{10}$ for $p=1/3$. We computed the values of $\gamma$ in Eq.~\eqref{gamma} for 25 different values of the parameter $x \in [0,0.08]$.  

For a fixed $x$ we expect the values of $\Gamma(x)$ to follow an approximately Gaussian distribution~\cite{Tutubalin-On-Limit-1965}, which allows us to obtain the average over the ensemble $\overline{\Gamma}(x)$ and its uncertainty $\Delta \overline{\Gamma}(x) = \sigma\{\Gamma\}/\sqrt{N_r}$, where $\sigma$ is the standard deviation. However, if we compare $\overline{\Gamma}(x)$ for two very close values of $x$, we observe significant fluctuations. We believe that this issue arises from a finite numerical precision in the components of the constituent $s$ matrix ({\color{Green}Eq.\ref{s-matrix}}). This matrix recurs in calculations, leading to error accumulation. The possibility of this issue being an artifact of the use of a pseudo-random number generator is ruled out, as it occurs even in a fixed disorder realization.

\begin{figure}[t]
\includegraphics[width=\linewidth]{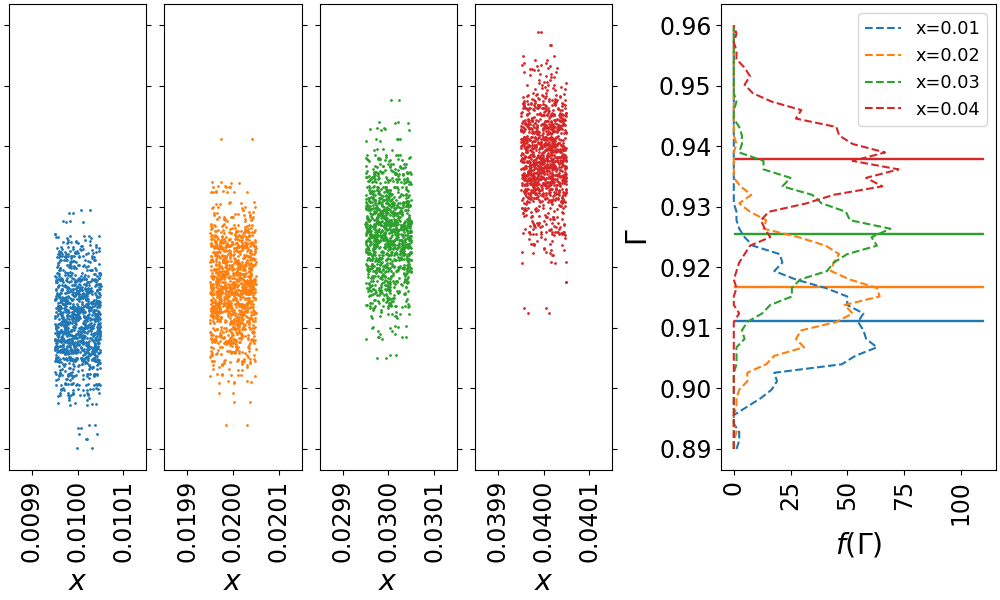}
\caption{Left: 
values of the rescaled Lyapunov exponent $\Gamma = M\gamma$ (for $M=120$) versus $x$ 
in small intervals centered at the points $x = 0.01$, $0.02$, $0.03$, and $0.04$,
Right: distributions of Lyapunov exponents in the clouds around these points, the solid lines show the averages with their thickness being the uncertainty.}
\label{fig:cloud}
\end{figure}

\begin{figure}[t]
\centering
\includegraphics[width=.49\linewidth]{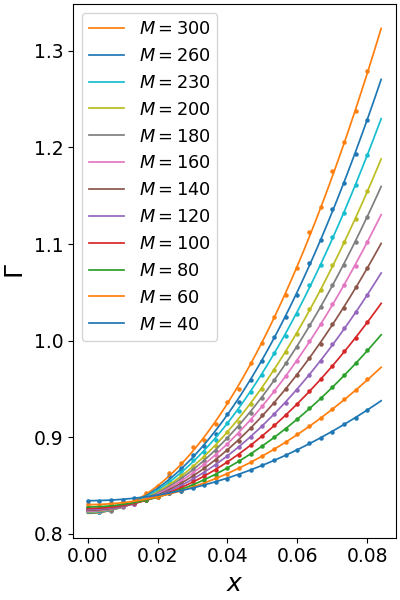}
\hfill
\includegraphics[width=.49\linewidth]{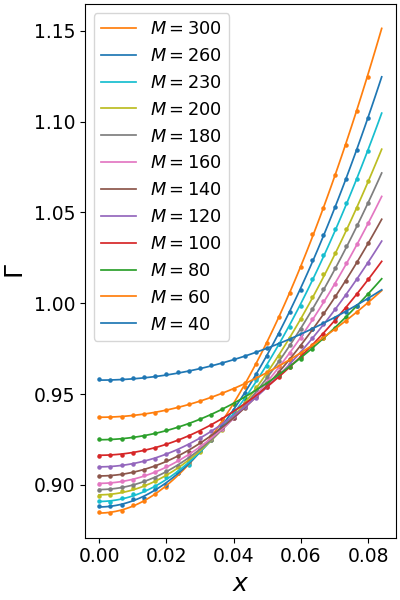}
\caption{The Lyapunov exponents $\Gamma$ for various system sizes $M$ and $x$ values for the CC model with $p=0$ (left) and RN model with $p=1/3$ (right), with the corresponding fitting curves.}
\label{fig:Lyapunov}
\end{figure}

To address this issue, we introduce a small window of width $2 \Delta x = 10^{-4}$ around each of the 25 ``nominal'' values of $x_s$, and sample $\Gamma(x)$ with $x$ randomly chosen from the range $[x_s - \Delta x, x_s + \Delta x]$. The resulting ensembles are also approximately Gaussian (see Fig.~\ref{fig:cloud}), allowing us to use the $x$-averaged values $\overline{\Gamma}_a(x_s)$. This procedure introduces an additional standard deviation $\sigma_a\{\Gamma\} = \overline{\Gamma}_a'(x_s) \cdot \Delta x/\sqrt{3}$ and a regular error $\Delta_a \overline{\Gamma}_a = \overline{\Gamma}_a'' \cdot {\Delta x}^2/6$. These quantities are negligibly small compared to the existing analogues: $\sigma_a\{\Gamma\} / \sigma\{\Gamma\} \lesssim 10^{-2}$ and $\Delta_a \overline{\Gamma}_a / \Delta \overline{\Gamma}_a \lesssim 10^{-4}$. The resulting data are presented in Fig.~\ref{fig:Lyapunov} together with the fitting curves.  


{\it Conclusions.} In this paper, we revisited random networks introduced and studied in Refs.~\cite{Gruzberg-Geometrically-2017, Kluemper-Random-2019}. These random networks have a parameter $p \in [0, 1/2]$ and reduce to the CC model for the integer quantum Hall transition at $p=0$. Our results demonstrate that the randomness of the network is a relevant disorder which changes the localization length exponent $\nu$ from its CC value. 

In Refs.~\cite{Gruzberg-Geometrically-2017, Kluemper-Random-2019} we employed the transfer matrix approach that required an ad-hoc regularizing parameter $\epsilon$ for open nodes. Here we introduce a novel approach that uses non-singular scattering matrices and avoids the need for $\epsilon$. In addition, the new approach avoids the appearance of exponentially large matrix elements in transfer matrices, and leads to a significant speed-up of numerical simulations. As a result, we are able to probe record network sizes, a larger number of the values of the parameter $x$, and larger ensembles of disorder realizations than in previous works.  

Our results for $p=0$ agree with previous findings~\cite{Slevin-Critical-2009, Obuse-Conformal-2010, Amado-Numerical-2011, Dahlhaus-Quantum-2011, Fulga-Topological-2011, Obuse-Finite-2012, Slevin-Finite-2012, Nuding-Localization-2015, Ippoliti-Integer-2018, Zhu-Localization-length-2019, Puschmann-Integer-2019, Kluemper-Random-2019, Ippoliti-Dimensional-2020, Puschmann-Edge-state-2021, Puschmann-Green's-2021, Sbierski-Criticality-2021, Dresselhaus-Numerical-2021, Huang-Numerical-2021, Dresselhaus-Scaling-2022, Bera-Quantum-2024}. For $p=1/3$ we confirm our previous results~\cite{Gruzberg-Geometrically-2017, Kluemper-Random-2019} including the value $\nu \approx 2.4$ for the critical exponent of the localization length, which is close to $\nu_{\text{exp}} \approx 2.38$ observed in experiments on the integer quantum Hall transition.

\textit{Acknowledgements.} We are grateful to F. Evers and T. Hakobyan for helpful discussions. The research was supported by Armenian SCS grants Nos. 20TTAT-QTa009 (HT, AS), 21AG-1C024 (HT, AS). AK acknowledges funding from the Deutsche Forschungsgemeinschaft (DFG) via Research Unit FOR 2316.
 
\bibliography{S-matrix-bibliography}

\end{document}